\documentclass[12pt]{scrartcl}

\setkomafont{disposition}{\normalfont\bfseries}

\usepackage{abstract}
\usepackage{authblk}
\usepackage{fullpage}
\usepackage{graphicx}
\usepackage[english]{babel}
\usepackage{color}
\usepackage[usenames]{xcolor}
\usepackage{amsfonts}
\usepackage{indentfirst}
\usepackage{setspace}
\usepackage{verbatim}
\usepackage{upgreek}
\usepackage{mathrsfs}
\usepackage{amsmath}
\usepackage{amsthm}
\usepackage{amssymb}
\usepackage{dsfont}
\usepackage{natbib}
\usepackage{float}
\usepackage{latexsym}

\raggedright
\newcommand{\IE}{i.e., }
\newcommand{\EG}{e.g., }

\newcommand{\IID}{{\it iid} }

\DeclareMathAlphabet{\msfsl}{T1}{cmr}{m}{it}
\DeclareMathAlphabet{\msyf}{OMX}{pcr}{m}{it}

\newcommand{\levelone}[1]{%
\section{#1}}

\newcommand{\leveltwo}[1]{%
\subsection{#1}}

\newcommand{\levelthree}[1]{%
\subsubsection{#1}}

\usepackage{xpatch}

\newlength{\myspace}
\makeatletter
\xpatchcmd{\@maketitle}{\vskip.5em}{\vskip\myspace}{}{}
\makeatother

\begin{document}

\title{The time-dependent reconstructed evolutionary process with a key-role for mass-extinction events}
\subtitle{\sc Research Article\\RH: The Reconstructed Process with Mass-Extinctions}
\author[1]{Sebastian H\"ohna}
\affil[1]{Department of Mathematics, Stockholm University, SE-106 91 Stockholm, Sweden}

\maketitle

\bigskip
\noindent{\bf Corresponding author:} Sebastian  H\"ohna, Department of Mathematics, Stockholm University, Stockholm, SE-106 91 Stockholm, Sweden; E-mail: Sebastian.Hoehna@gmail.com.\\

\newpage

\begin{abstract}
The homogeneous reconstructed evolutionary process is a birth-death process without observed extinct lineages. 
Each species evolves independently with the same diversification rate---speciation rate, $\lambda(t)$, and extinction rate, $\mu(t)$---that may change over time. 
The process is commonly applied to model species diversification where the data are reconstructed phylogenies, \EG trees estimated from present-day molecular data, and used to infer diversification rates.

In the present paper I develop the general probability density of a reconstructed tree under any homogeneous, time-dependent birth-death process. 
I demonstrate how to adapt this probability density when conditioning on the survival of one or two initial lineages, or on the process realizing $n$ species, and also how to transform between the probability density of a reconstructed tree and the probability density of the speciation times.

I demonstrate the use of the general time-dependent probability density functions by deriving the probability density of a reconstructed tree under a birth-death-shift model with explicit mass-extinction events. 
I extend these functions to several special cases, including the pure-birth process, the pure-death process, the birth-death process, and the critical-branching process. 
Thus, I specify equations for the most commonly used birth-death models in a unified framework (\EG same condition and same data) using a common notation.

\end{abstract}

\textbf{\textsc{key words:}} Birth-Death Process, Speciation, Diversification, Mass-Extinction, Incomplete Taxon Sampling, Probability Density Function, Likelihood \\
\newpage
\levelone{Introduction}
The birth-death process is commonly used to model species diversification and to infer diversification rates (speciation and extinction rates) from reconstructed phylogenies \citep{Nee2006}. 
Likelihood-based estimates, whether based on a maximum likelihood or Bayesian framework, require the probability density function of the reconstructed tree under a birth-death process and dominate parameter estimation methods in phylogenetics \citep{Huelsenbeck2001,Holder2003}. 
Besides their use in parameter estimation, the probability density functions are crucial for hypothesis testing, \EG in testing whether rates have been constant or variable over time \citep{Huelsenbeck1997,Rabosky2006}.

The probability density function of a reconstructed tree under the reconstructed evolutionary process has been derived under various scenarios by different authors \citep{Nee1994,Rabosky2006,Morlon2011,Stadler2011,Etienne2012} and has been applied in several studies (for reviews see \cite{Ricklefs2004}, \cite{Nee2006}, \cite{Ricklefs2007} and \cite{Pyron2013}). 
However, it remains challenging to compare these probability density functions because they differ in their notation, derivation and conditioning, \EG conditioning on survival of the process or conditioning on obtaining exactly $n$ species \citep{Stadler2013}. 
Furthermore, the probability density functions are inconsistently applied to reconstructed trees or speciation times: each requires different combinatorial factors.
This inconsistency prevents the use of model-selection methods and so precludes the comparison of candidate models.

The present paper provides a thorough study of the time-dependent homogeneous reconstructed evolutionary process under various time-dependent diversification rate functions and serves as a compendium of probability distribution functions presented in a common notation. 
I start by deriving the probability density of a reconstructed tree under the time-dependent birth-death process in the general case, \IE with any diversification rate functions. 
I then demonstrate how to condition on the survival of one initial lineage (where the process starts at the stem node of the tree), two initial lineages (where the process starts at the crown node of the tree), or on the process realizing $n$ species today.
Additionally I show how any of the derived probability densities can be transformed to apply for reconstructed trees or the speciation times of a reconstructed tree only.

The utility of the probability densities presented here lies in their applicability to any diversification rate functions.
I demonstrate this flexibility by deriving the explicit probability density of a reconstructed tree under a birth-death-shift model, \IE piecewise constant diversification rates, with explicit mass-extinction events.
I complete this discussion on the time-dependent reconstructed evolutionary process with mass-extinction events by providing the probability densities of a reconstructed tree under a pure-birth process with constant-rate and exponentially decaying rate and a constant-rate birth-death process.

\begin{figure*}[!tpb]
       \centerline{\includegraphics[width=\textwidth]{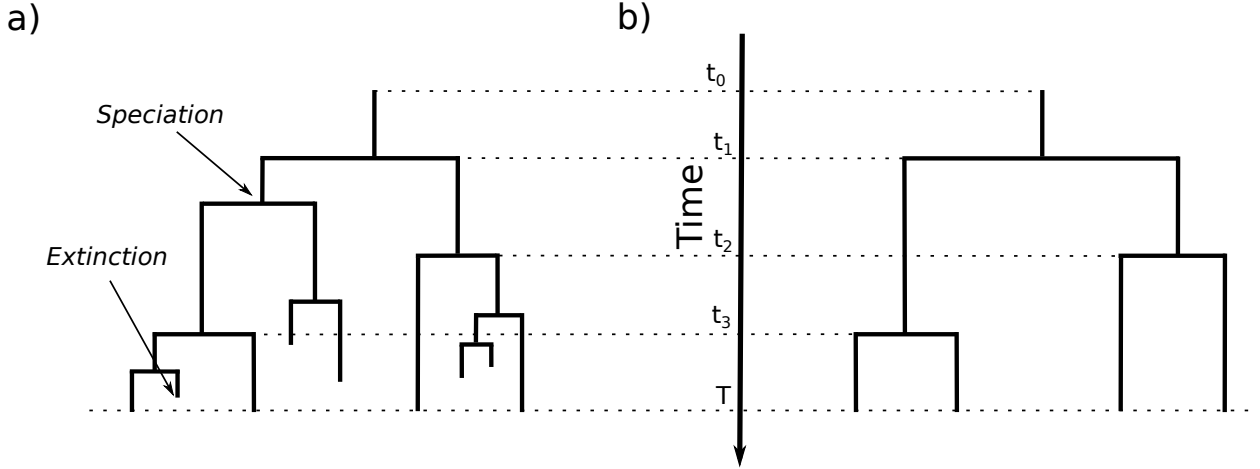}}
   \caption{A simulated birth-death tree starting with a single species at time $t_0$, thus $N(t_0)=1$. The process was stopped at time $T$. a) The complete tree containing both extant and extinct species. b) The reconstructed tree containing only extant species. Speciation events in the reconstructed tree occurred at times $t_1$, $t_2$ and $t_3$.}\label{fig:01}
\end{figure*}

\levelone{The reconstructed evolutionary process}
I define the birth-death process with non-constant rates for rooted, strictly bifurcating trees following the notation of \cite{Nee1994}. 
Let $N(t)$ denote the number of species alive at time $t$. 
Furthermore, let the process start with a single species at time $t_0$, such that $N(t_0) = 1$. 
A speciation event increases the number of species by one---\IE from $k$ to $k+1$ assuming that $k$ species are alive at time $t$---after an exponentially distributed time with rate $k\lambda(t)$.
Similarly, an extinction event decreases the number of species by one after an exponentially distributed time with rate $k\mu(t)$. 
At a speciation event, one of the $k$ species is replaced by two new descendant species, where the probability of each species giving birth is equally probable. 
At an extinction event, one species simply dies, where each species has the same probability of going extinct. 
Commonly, the process is stopped at the present time, denoted $T$, and the number of extant species is denoted $N(T) = n$. 
Figure~\ref{fig:01}a depicts a binary tree resulting from a birth-death process, showing both extant and extinct lineages. 
This is known as a \emph{complete tree}. 
Figure~\ref{fig:01}b shows the same tree but after removing all extinct lineages; this is a \emph{reconstructed tree}. 
The reconstructed trees are the data (observations) that I consider here.

\cite{Kendall1948} derived the probability that a process survives ($N(T) > 0$) and the probability of obtaining exactly $n$ species at time $T$ ($N(T) = n$) when the process started at time $t$ with one species. 
Kendall's results were summarized in Equation (3) and Equation (24) of \cite{Nee1994}
\begin{eqnarray}
P(N(T)\!>\!0|N(t)\!=\!1) & = & \left(1+\int\limits_t^{T} \bigg(\mu(s) \exp(r(t,s))\bigg) ds\right)^{-1} \label{eq:survival} \\
P(N(T)\!=\!n|N(t)\!=\!1) & = & (1-P(N(T)\!>\!0|N(t)\!=\!1)\exp(r(t,T)))^{n-1} \nonumber\\
& & \times P(N(T)\!>\!0|N(t)\!=\!1)^2 \exp(r(t,T)) \label{eq:N} 
\end{eqnarray}
where $r(t,s) = \int_t^s\mu(x)-\lambda(x) dx$. 
Note that the notation of $r(t,s)$ follows the notation of \cite{Nee1994} but one could also use the perhaps more intuitive form $r(t,s) = \int_t^s\lambda(x)-\mu(x) dx$ and replace each occurrence of $r(t,s)$ by $-r(t,s)$.

These two probability functions are sufficient to specify the probability density of a reconstructed tree, as I will show below.
Thus, an analytical solution for the probability density of reconstructed trees requires only that we have an analytical solution for the rate integral $r(t,s)$ and the probability of survival $P(N(T)\!>\!0|N(t)\!=\!1)$.

Note that the probability of $n$ extant species, conditioned on survival of the process, is geometrically distributed with parameter $p=P(N(t)\!>\!0|N(t_0)\!=\!1)\exp(r(t_0,t))$. 
Therefore, the expected number of species alive at time $t$ is given by
\begin{eqnarray}
E[N(t) | S(1,t_0,t)] & = & \big(P(N(t)\!>\!0|N(t_0)\!=\!1)\exp(r(t_0,t))\big)^{-1}
\end{eqnarray}
where $S(i,t_0,t)$ denotes that the $i$ species alive at time $t_0$ each have at least one descendant at time $t$.
The expected number of species not conditioning on survival is obtained by splitting the equation into the two scenarios: when the process results in extinction and when the process results in survival of the original lineage.
\begin{eqnarray}
E[N(t)] & = & \big(1-P(N(t)\!>\!0|N(t_0)\!=\!1)\big) \times 0 \nonumber \\
& & + P(N(t)\!>\!0|N(t_0)\!=\!1) \times \big(P(N(t)\!>\!0|N(t_0)\!=\!1)\exp(r(t_0,t))\big)^{-1} \nonumber \\
& = & \frac{P(N(t)\!>\!0|N(t_0)\!=\!1)}{P(N(t)\!>\!0|N(t_0)\!=\!1)\exp(r(t_0,t))} \nonumber  \\
& = & \big(\exp(r(t_0,t))\big)^{-1} \mbox{ .}
\end{eqnarray}

\leveltwo{Probability density of a reconstructed evolutionary tree}
Let $\Psi$ denote a reconstructed evolutionary tree comprising a tree topology $\tau$ and the set of branching times $\mathbb{T}$. 
The probability density of all speciation times $\mathbb{T}$ in the reconstructed tree is
\begin{eqnarray}
f(\mathbb{T}|N(t_0)\!=\!1)  & = & P(N(T)\!=\!1|N(t_0)\!=\!1) \times P(N(T)\!>\!0|N(t_0)\!=\!1) \nonumber\\
& & \times\prod_{i=1}^{n-1}\big(i\times\lambda(t_i)\times P(N(T)\!=\!1|N(t_i)\!=\!1)\big) \nonumber \\
  & = & P(N(T)\!>\!0|N(t_0)\!=\!1)^3 \exp(r(t_0,T))  \nonumber\\
& & \times\prod_{i=1}^{n-1}\big(i\times\lambda(t_i)\times P(N(T)\!>\!0|N(t_i)\!=\!1)^2 \exp(r(t_i,T))\big)  \label{eq:times}
\end{eqnarray}
which was derived by \citet[Equation (3.4.6)]{Thompson1975} for constant rates (see also Equation~(20) in \cite{Nee1994}). A short proof of this equation for arbitrary rates was given in the Appendix in \cite{Hohna2013a}.

One way of understanding the probability density of a reconstructed tree is to view it as the result of multiplying together the probability densities of the speciation events ($\lambda(t_i)$) and the probability densities of obtaining exactly one species originating from each speciation event ($P(N(T)\!=\!1|N(t_i)\!=\!1)$).
We can ignore every other speciation event along the branches that must lead to an extinct species.

Equation~(\ref{eq:times}) holds for any homogeneous, time-dependent birth-death process. In the later sections I will insert the rate specific probability densities to obtain the probability density of $\mathbb{T}$.

\levelthree{Converting between the probability of a reconstructed tree and probability of the speciation times}
The previous probability density is valid for set of speciation times $\mathbb{T}$. 
In other scenarios, for example when used as priors in Bayesian phylogenetic inference, the probability density of a reconstructed tree is needed.
To be precise, the birth-death process specifies a distribution on labeled histories instead of tree topologies and thus I interpret reconstructed trees as labeled histories.

There are $n!(n-1)!/2^{n-1}$ distinct labeled histories; each is equiprobable under a birth-death process \citep{Edwards1970,Rannala1996}.
Hence, the probability density of the reconstructed tree $\Psi$ is given by
\begin{equation}
f(\Psi) = \frac{2^{n-1}}{n!(n-1)!}f(\mathbb{T}) \label{eq:conversion}
\end{equation}
see Equation~(4) in \cite{Rannala1996}. 
Any of the following probability density functions will be obtained for the set of speciation times but can be converted using Equation~(\ref{eq:conversion}).

\levelthree{Conditioning on survival of the process}
It is often argued that the birth-death process should be conditioned on survival because otherwise we could not have observed a tree \citep{Nee1994}. 
Thus, the probability density of the speciation times needs to be divided by Equation~(\ref{eq:survival})
\begin{eqnarray}
f(\mathbb{T}|N(t_0)\!=\!1,S(1,t_0,T))  & = & f(\mathbb{T}|N(t_0)\!=\!1) / P(N(T)\!>\!0|N(t_0)\!=\!1) \nonumber \\
& = & P(N(T)\!=\!1|N(t_0)\!=\!1) \nonumber\\
& & \times\prod_{i=1}^{n-1}\big(i\times\lambda(t_i)\times P(N(T)\!=\!1|N(t_i)\!=\!1)\big) \label{eq:timesSurvival}
\end{eqnarray}
where $S(i,t_0,T)$ again denotes that the $i$ lineages alive at time $t_0$ leave at least one descendant at time $T$.

Conditioning on survival has a peculiar side-effect that is discussed later in the context of a critical branching process.

\levelthree{Conditioning on the number of observed species}
\cite{Rannala1996} argued further that one should condition on the number of extant species because the number of species is fixed in any common phylogenetic analysis. Thus, $f(\mathbb{T})$ is divided by the probability of observing $n$ species, given in Equation~(\ref{eq:N}), which yields
\begin{eqnarray}
f(\mathbb{T}|N(t_0)\!=\!1,N(T)\!=\!n)  & = & f(\mathbb{T}|N(t_0)\!=\!1) / P(N(T)\!=\!n|N(t_0)\!=\!1) \nonumber \\
 & = & \prod_{i=1}^{n-1}\left(\frac{i\times\lambda(t_i)\times P(N(T)\!=\!1|N(t_i)\!=\!1)}{1-P(N(T)\!>\!0|N(t_0)\!=\!1)\exp(r(t_0,T))}\right) \mbox{ .} \label{eq:timesN}
\end{eqnarray}

\cite{Gernhard2008} argued that one should only condition on the the number of observed species because no information about the time since the origin of the process is known.
Unfortunately, in order to condition only on the number of extant species \cite{Gernhard2008} implicitly assumed a uniform prior on the time of the process. 
Furthermore, analytical solutions to the probability density have only been obtained using a uniform prior for the time of the origin and constant-rate functions. 
Therefore I will not pursue the issue further. 
I have nevertheless included equations for conditioning on both the time and the number of extant species for the sake of completeness.

Conditioning on the number of observed species is reasonable only if diversification rates are estimated for multiple phylogenies each containing the number of species, \EG what are the estimated diversification rates for phylogenies with 100 species.
However, I think it is unlikely that a single reconstructed tree was obtained by collecting data and throwing away all groups that did not have exactly $n$ species. 
Therefore, in cases when only one phylogeny is considered one should not condition on the number of extant species. 
The number of species is part of the observation and not fixed before seeing the data.
Nevertheless, conditioning on the number of observed taxa and the age of the tree is mathematically convenient, as each speciation time in the reconstructed tree is independent and identically distributed (\IID) \citep{Rannala1996,Lambert2010,Hohna2011,Hohna2013b}.


\levelthree{Starting at the most recent common ancestor}
Most phylogenetic analyses lack information regarding the length of the root branch (\IE the `stem' age of tree). 
Instead, analyses provide information on the age of the \emph{most recent common ancestor} (MRCA, \IE the `crown' age of the tree), which corresponds to $t_{MRCA} = t_1$. 
Thus, the process starts at time $t_1$ instead of time $t_0$ and from two instead of one species. 
Furthermore, it is necessary here to condition on survival of both initial lineages because the extinction of one of the lineages results in a different $t_{MRCA}$ for the tree. 
The probability density of the speciation times is given by
\begin{eqnarray}
f(\mathbb{T}|N(t_1)\!=\!2,S(2,t_1,T))  & = & \bigg(P(N(T)\!=\!1|N(t_1)\!=\!1)\bigg)^2 \nonumber\\
& & \times\prod_{i=2}^{n-1}\big(i\times\lambda(t_i)\times P(N(T)\!=\!1|N(t_i)\!=\!1)\big) \mbox{ .} \label{eq:timesMRCA}
\end{eqnarray}

Furthermore, from \cite{Rannala1996} we have the probability function on $n$ species conditioned on starting at the MRCA for constant rates and from \cite{Hohna2013a} under time-dependent rates given by
\begin{eqnarray}
\lefteqn{P(N(T)\!=\!n|N(t_1)\!=\!2,S(2,t_1,T))} \nonumber\\
& = & \sum_{i=1}^{n-1}\Big(P(N(T)\!=\!i|N(t_1)\!=\!1,S(1,t_1,T)) \times \nonumber \\
& & P(N(T)\!=\!n-i|N(t_1)\!=\!1,S(1,t_1,T))\Big) \nonumber \\
& = & (n-1) (P(N(T)\!>\!0|N(t_1)\!=\!1)^2 \exp(r(t_1,T)))^2 \nonumber\\
& & \times (1-P(N(T)\!>\!0|N(t_1)\!=\!1)\exp(r(t_1,T)))^{n-2} \mbox{ .} \label{eq:Nmrca}
\end{eqnarray}


\leveltwo{Probability density and distribution function of a speciation event in the reconstructed tree}
The time of a speciation event in a reconstructed tree conditioned on the age of the tree and the number of extant taxa is \IID \citep{Rannala1996,Lambert2010,Hohna2011,Hohna2013b}. 
This fact can be exploited to provide efficient simulation of reconstructed trees \citep{Hohna2013a}, to integrate over possible times a missing speciation event could have happened \citep{Hohna2011,Hohna2013b}, and to condition on a known speciation event \citep{Yang2006}.
The probability density function of the divergence times for non-constant rates given that the speciation event happened is (Equation~5 in \cite{Hohna2013b})
\begin{equation}
f(t|t_0\leq t \leq T) = \frac{\lambda(t)P(N(T)\!=\!1|N(t)\!=\!1)}{1-P(N(T)>0|N(t_0)=1)\exp{(r(t_0,T))}} \label{eq:f_iid}
\end{equation}
and the distribution function is (Equation~6 in \cite{Hohna2013b})
\begin{equation}
F(t|N(t_0)=1,t_0\leq t \leq T) = 1 - \frac{1-P(N(T)>0|N(t)=1)\exp{(r(t,T))}}{1-P(N(T)>0|N(t_0)=1)\exp{(r(t_0,T))}}  \mbox{ .}\label{spec_dist}
\end{equation}

\begin{figure*}[!tpb]
       \centerline{\includegraphics[width=0.5\textwidth]{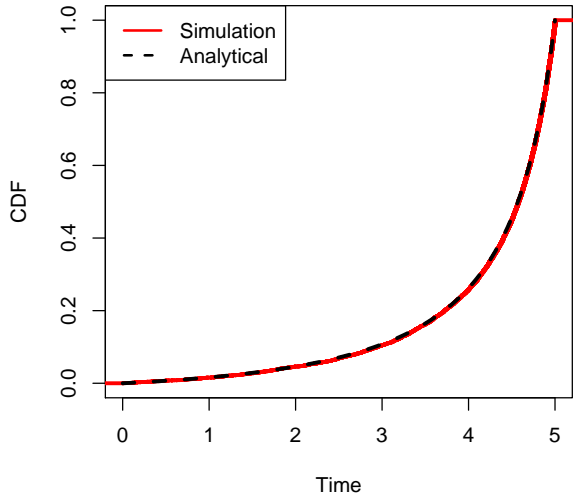}}
   \caption{Empirical (simulated) and analytical probability distribution function of the time of a speciation event in the reconstructed tree. The process was simulated with diversification rates $\lambda = 2.0$ and $\mu = 1.5$ and was stopped at $T = 5.0$. }\label{fig:03}
\end{figure*}

The probability density of the speciation times in a reconstruction tree can be obtained from Equation~(\ref{eq:timesN}) (see also \cite{Hohna2013a,Hohna2013b}). 
Furthermore, \cite{Hohna2013b} derived the corresponding distribution function. 
Nevertheless, I validated Equation~(\ref{spec_dist}) by simulating birth-death trees, pruning away all extinct lineages and thus recording the speciation times of the reconstructed trees. 
The simulated distribution of speciation times matches the analytically derived distribution function, see Figure~\ref{fig:03}.

\begin{figure*}[!tpb]
       \centerline{\includegraphics[width=\textwidth]{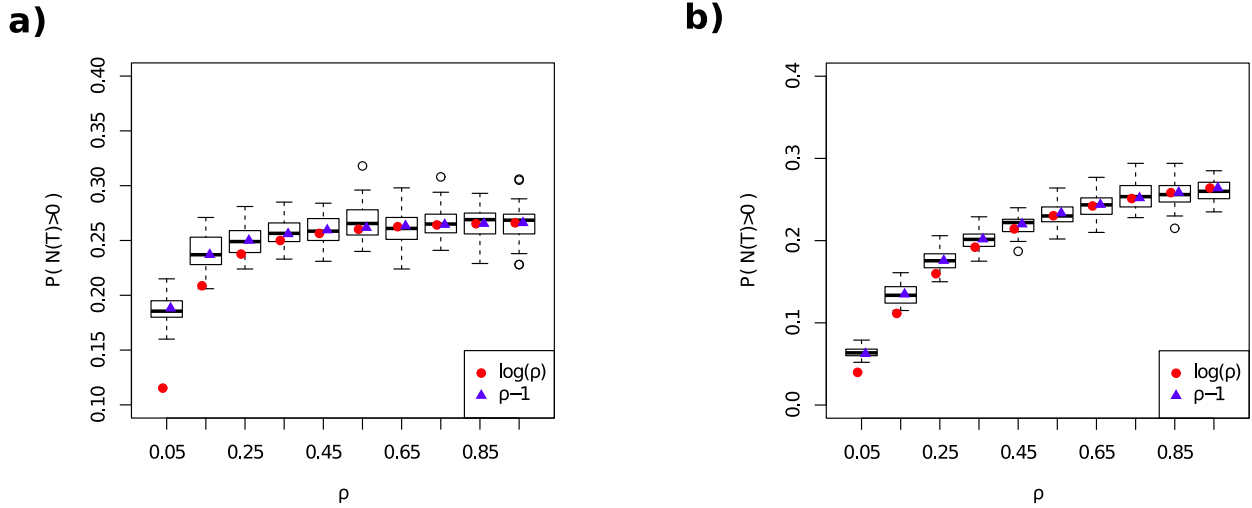}}
   \caption{The probability that the birth-death process survives when a) species are included with probability $\rho$, or b) when a mass-extinction event occurs. A mass-extinction event occurred at time $t_m = 3.0$, and each species survived the event with probability $\rho$. The process was simulated with diversification rates $\lambda = 2.0$ and $\mu = 1.5$ and terminated at $T = 5.0$. The box-plots show the simulated survival probabilities of 50 repetitions (data points) of each 1000 simulation. Additionally, the analytical solution for the survival probability is given with and without the approximation $\ln(\rho) = \rho - 1$.}\label{fig:04}
\end{figure*}

\leveltwo{Probability density of a reconstructed tree with uniform taxon sampling}
The reconstructed tree is often incomplete because several species are missing. 
The standard approach is to accommodate species sampling---or taxon sampling---by assuming that each species is sampled with uniform probability, \IE each species is included with probability $\rho$ \citep{Nee1994,Yang1997,Stadler2009,Hohna2011}. 
\cite{Nee1994} suggested that uniform taxon sampling can be modeled using
\begin{equation}
\int_t^s\mu(x) dx = \begin{cases}
-\ln(\rho) + \int_t^s \mu'(x) dx & \text{ if }t < T \leq s\\
\int_t^s \mu'(x) dx & \text{ otherwise}
\end{cases} \label{eq:extinctionRateSampling}
\end{equation}
where $\mu'(x)$ is the extinction rate with \emph{complete} taxon sampling \citet[see Equation~(31)]{Nee1994}. 
Then, the diversification rate integral is
\begin{equation}
r(t,s) = \begin{cases}
-\ln(\rho) + \int_t^s \big(\mu'(x)-\lambda(x)\big) dx & \text{ if }t < T \leq s\\
\int_t^s \big(\mu'(x)-\lambda(x)\big) dx & \text{ otherwise.}
\end{cases}
\end{equation}
The probability that the process of survives---or that at least one species is sampled---is computed by
\begin{eqnarray}
\lefteqn{P(N(T)\!>\!0|N(t)\!=\!1)} \nonumber\\
& = & \left(1+\int\limits_t^{T} \bigg(\mu(s) \exp(r(t,s))\bigg) ds\right)^{-1} \nonumber\\
& = & \left(1+\int\limits_t^{T-\Delta t} \bigg(\mu'(s) \exp(r'(t,s))\bigg) ds + \int\limits_{T-\Delta t}^{T} \bigg(\mu(s) \exp(r(t,s))\bigg) ds\right)^{-1} \nonumber \\
& \stackrel{\Delta t \to 0}{=} & \left(1+\int\limits_t^{T} \bigg(\mu'(s) \exp(r'(t,s))\bigg) ds - \ln(\rho) \exp(r'(t,T)-\ln(\rho))\right)^{-1} \nonumber \\
& \approx & \left(1+\int\limits_t^{T} \bigg(\mu'(s) \exp(r'(t,s))\bigg) ds - \frac{\rho-1}{\rho} \exp(r'(t,T))\right)^{-1} \mbox{ .} \label{eq:uniformSampling}
\end{eqnarray}
Equation~(\ref{eq:uniformSampling}) is derived by splitting the integral into the interval that excludes the sampling time $[t_0,T-\Delta t]$ and the interval that includes the sampling time $(T-\Delta t,T]$. 
Then, the derivation uses the fact that $\int_{T-\Delta t}^{T}\mu(x) dx = -\ln(\rho)$ if $\Delta t$ converges to zero. 
The last step of Equation~(\ref{eq:uniformSampling}) is reached by substituting $\ln(\rho)$ with $\rho-1$. 
This substitution may appear to be a crude approximation. 
However, the resulting equation is equivalent to Equation~(34) in \cite{Nee1994} (see also \cite{Yang1997} and \cite{Lambert2010}). 
Thus, I performed simulations to explore whether this approximation provides acceptable results, see Figure~\ref{fig:04}a. 
The substituted equation outperforms the original equation and seems to give very accurate results when compared to the simulations.
This observation indicates that using the extinction rate function given in Equation~\ref{eq:extinctionRateSampling} is only an approximation for actual uniform taxon sampling but with the correct substitution gives the correct probability of survival.

Note that the extinction rate, $\mu'(x)$, and the diversification rate integral, $r'(t,s)$, are equivalent to the original equation. 
Thus, Equation~\ref{eq:uniformSampling} differs from Equation~(\ref{eq:survival}) only by the term $\frac{\rho-1}{\rho} \exp(r'(t,s))$. 
This is the only term that needs to be included when incorporating uniform taxon sampling under any homogeneous birth-death model.
Inserting $\rho=1$ confirms that this equation simplifies to the complete-sampling approach used above.

\leveltwo{Including a single mass-extinction event}
Mass-extinction events can be modeled using the same approach as uniform taxon sampling. 
The uniform taxon sampling model specifies that each species is sampled at the present with the same probability, $\rho$.
Similarly, we can model a mass-extinction event by assuming that every species alive at the time of the mass-extinction event, $t_m$, survives with the same probability, $\rho_m$. 
Thus, the extinction rate under a model with explicit mass-extinction events is
\begin{equation}
\int_t^s\mu(x) dx = \begin{cases}
-\ln(\rho_m) + \int_t^s \mu'(x) dx & \text{ if }t < t_m \leq s\\
\int_t^s \mu'(x) dx & \text{ otherwise}
\end{cases}
\end{equation}
where $\mu'(x)$ again denotes the original extinction rate (without the mass-extinction event). 
Then, the diversification rate integral is
\begin{equation}
r(t,s) = \begin{cases}
-\ln(\rho_m) + \int_t^s \big(\mu'(x)-\lambda(x)\big) dx & \text{ if }t < t_m \leq s\\
\int_t^s \big(\mu'(x)-\lambda(x)\big) dx & \text{ otherwise}
\end{cases}
\end{equation}
and the original diversification rate integral is $r'(s,t) = \int_t^s \big(\mu'(x)-\lambda(x)\big) dx$.

The probability that the process survives given the mass-extinction event is
\begin{eqnarray}
\lefteqn{P(N(T)\!>\!0|N(t)\!=\!1)} \nonumber\\
& = & \left(1+\int\limits_t^{T} \bigg(\mu(s) \exp(r(t,s))\bigg) ds\right)^{-1} \nonumber\\
& = & \Bigg(1+\int\limits_t^{t_m-\Delta t} \bigg(\mu'(s) \exp(r(t,s))\bigg) ds + \int\limits_{t_m-\Delta t}^{t_m+\Delta t} \bigg(\mu(s) \exp(r(t,s))\bigg) ds \nonumber \\
&    & +\int\limits_{t_m+\Delta t}^{T} \bigg(\mu'(s) \exp(r(t,s))\bigg) ds\Bigg)^{-1} \nonumber \\
& \stackrel{\Delta t \to 0}{=} & \left(1+\int\limits_t^{T} \bigg(\mu'(s) \exp(r(t,s))\bigg) ds - \ln(\rho) \exp(r'(t,t_m)-\ln(\rho_m))\right)^{-1} \nonumber \\
& \approx & \left(1+\int\limits_t^{T} \bigg(\mu'(s) \exp(r(t,s))\bigg) ds - \frac{\rho_m-1}{\rho_m} \exp(r'(t,t_m))\right)^{-1} \mbox{ .} \label{eq:massExtinction}
\end{eqnarray}
Equation~(\ref{eq:massExtinction}) is derived again by splitting the integral into different intervals---intervals that include the mass-extinction event and intervals that do not---and letting the size of the interval that contains the mass-extinction event go to zero. 
Furthermore, $\ln(\rho_m)$ is again substituted with $\rho_m - 1$ in the last step, and the adequacy of this approximation investigated by simulation. 
Figure~\ref{fig:04}b indicates that the approximation provides a very accurate expression for the survival probability.

Inserting $t_m = T$ in Equation~(\ref{eq:massExtinction}) simplifies to the probability of survival under uniform taxon sampling (Equation~(\ref{eq:uniformSampling})), and inserting $\rho = 1.0$ further simplifies the equation to that for the survival probability without a mass-extinction event (Equation~(\ref{eq:survival})).

\leveltwo{Including multiple mass-extinction events}
Now, consider the case when $k$ mass-extinction events have occurred at the times $\mathbb{M} = \{m_1,\ldots,m_k\}$ each with a different survival probability,  denoted  $\mathbb{P} = \{\rho_1,\ldots,\rho_k\}$. 
Let us assume, without loss of generality, that some mass-extinction events occurred between the start of the process, $t$, and the present, $T$, so that $t < m_i < \ldots < m_j \leq T$. 
Then, the probability that the process survives is given by
\begin{eqnarray}
\lefteqn{P(N(T)\!>\!0|N(t)\!=\!1)} \nonumber\\
& = & \left(1+\int\limits_t^{T} \bigg(\mu'(s) \exp(r(t,s))\bigg) ds - \sum_{k = i}^j\frac{\rho_k-1}{\prod_{l=i}^k\rho_l} \exp(r'(t,m_k))\right)^{-1} \mbox{ .} \label{eq:massExtinction}
\end{eqnarray}
In \cite{Hohna2013a} Equation~(\ref{eq:massExtinction}) was originally derived with a slightly different representation, $\sum\limits_{k = i}^{j}(\rho_k - 1)\exp(r(t,m_k))$, which is exactly equivalent because $\exp(r(t,m_k)) = \frac{1}{\prod_{k=i}^j \rho_k} \exp(r'(t,m_k))$. 
The only difference stems from the use of $r(t,s)$ or $r'(t,s)$ which contain $\mu(x)$ and $\mu'(x)$ respectively.

\levelone{The Birth-Death-Shift Process with Mass-Extinction Events}
The time-dependent birth-death process and the probability densities can be used with any diversification rate function. 
Unfortunately, numerical integration is necessary for most choices of the rate function with the exception of the constant-rate process discussed in the Appendix. 
As an example, I derive the probability density under the birth-death-shift process. 
Moreover, any constant-rate birth-death process, including the constant-rate pure-birth process, and the pure-birth-shift process can be considered as a special case of the birth-death-shift process.

The rate-shift model specifies that the diversification rates are constant over a given time interval and shift abruptly at specific times. 
The model was first applied with a single rate shift for the speciation rate \citep{Rabosky2006} and then extended to any number of rate shifts for both the speciation and extinction rate \citep{Stadler2011}. 
Here I provide the probability density for survival of at least one lineage following the derivation in the previous section. 
The explicit equation of this probability density, $P(N(T)\!>\!0|N(t)\!=\!1)$, is sufficient to compute the probability density of a reconstructed tree, see Equation~\ref{eq:times}.

Let the vector $\mathbb{S} = \{s_1,\ldots,s_k\}$ denote the times of the $k$ rate shifts. 
I assume, for notational convenience, that $s_0 = t_0$ represents the origin of the process. 
Furthermore, let the vector $\mathbb{B} = \{b_0,\ldots,b_k\}$ denote the speciation rate in the interval $t \in (s_i,s_{i+1}]$ and $\mathbb{D} = \{d_0,\ldots,d_k\}$ the extinction rate respectively.
Thus, the speciation rate function is $\lambda(t) = b_i$ for $s_i \leq t < s_{i+1}$ and the extinction rate function is $\mu(t) = d_i$ for $s_i \leq t < s_{i+1}$. 
As before $\rho_i$ denotes the probability of the process surviving a mass-extinction event at time $s_i$. 
For convenience of notation I assume that mass-extinction events only occur at rate-shift times, but one can imagine additional rate-shift times for the mass-extinction events if the diversification rates do not change at a mass-extinction event.

\makeatletter
\newcommand{\vast}{\bBigg@{4}}
\newcommand{\Vast}{\bBigg@{5}}
\makeatother

The probability density of survival of at least one lineage is given by
\begin{eqnarray}
\lefteqn{P(N(T)\!>\!0|N(t)\!=\!1)} \nonumber \\
 & = & \left(1+\int\limits_t^{T} \bigg(\mu(s) \exp(r(t,s))\bigg) ds\right)^{-1} \nonumber \\
 & = & \left(1+ \sum\limits_{i = 0}^{k} \left( \int\limits_{t_i}^{t_{i+1}-\Delta t} \bigg(\mu \times \text{e}^{r(t,s)}\bigg) ds + \int\limits_{t_{i+1}-\Delta t}^{t_{i+1}+\Delta t} \bigg(\mu(s) \text{e}^{r(t,s)}\bigg) ds \right) \right)^{-1} \nonumber \\
& \overset{\Delta t \to 0}{=} & \Vast(1+\sum\limits_{i = 0}^{k}\vast( \frac{d_i}{d_i - b_i} \times \text{e}^{\sum\limits_{j = 0}^{i-1}(d_j-b_j)(t_{j+1}-t_j)- \ln(\rho_j)} \times\bigg(\text{e}^{(d_i - b_i)(t_{i+1}-t_i)} - 1\bigg) \nonumber \\
& &  - \frac{\rho_i-1}{\prod_{j=1}^{i}\rho_j} \times \text{e}^{\sum\limits_{j = 0}^{i-1}(d_j-b_j)(t_{j+1}-t_j)} \vast)\Vast)^{-1} \mbox{ .}
\end{eqnarray}
The resulting probability density of a reconstructed tree is equivalent to the equation in \cite{Stadler2011}. 
However, the independent derivation here confirms the results by Stadler. 
My motivation for presenting the birth-death-shift process here are two-fold: first, demonstrating the flexibility of the general time-dependent rate functions for any specific instance, and, second, to present the birth-death-shift process within the same notation as all other process presented in the paper.

\levelone{Discussion}

\leveltwo{Parameterization and constraints}
The choice of parameterization and parameter constraints has an important impact of diversification-rate estimates.
Often, the resulting effect is unintended.
For example, a common prior belief is that the speciation rate is greater than the extinction rate.
Instead of using separate parameters for the speciation and extinction rates, a composite prior, such as $\frac{\mu}{\lambda} \sim \text{Uniform}(0,1)$, is often applied \citep[as implemented in BEAST,][]{Drummond2012}. 
However, combining this net-diversification rate prior with the relative-extinction prior, $\lambda - \mu \sim \text{Uniform}(a,b)$, induces equal prior probabilities on $\lambda$ between $a$ and $b$ ($P(\lambda) \propto \frac{1}{b-a}$) and a decreasing probability for $\lambda \ge b$ ($P(\lambda) \propto \frac{1}{\lambda}$). 
The induced prior probability on $\mu$ is strongly concentrated on small values ($P(\mu) \propto \frac{1}{b}\ln{(\frac{b+\mu}{\mu})}$). 
Note that all induced priors are improper priors which may cause problems in Bayesian model selection methods (see, \EG \cite{Baele2013}).

Moreover, the probability densities derived here do not require that $\lambda > \mu$, either that this condition holds at any instant or that $\int\lambda(x)dx > \int\mu(x)dx$. 
These constraints are necessarily violated under a model with mass-extinction events because the extinction rate far exceeds the speciation rate during the mass-extinction event. 
Furthermore, I will also show below that these probabilities hold for a pure-death process. 
Henceforth, I argue that it is not necessary to constrain $\lambda(t) > \mu(t)$.

\leveltwo{Pure-birth processes}

All pure-birth models have an extinction rate of $\mu(t) = 0$. 
The probability of survival must therefore be one because extinction cannot occur. 
This can also be seen by inserting the extinction rate ($\mu(t) = 0$) into the equation
\begin{eqnarray}
P(N(T)\!>\!0|N(t)\!=\!1) = \left(1+\int\limits_t^{T} \bigg(\mu(s) \exp(r(t,s))\bigg) ds\right)^{-1} = 1 \mbox{ .}\nonumber
\end{eqnarray}
Furthermore, conditioning on survival does not change the probability densities.

The fact that the probability of survival is always equal to one for a pure-birth process means that analytical probability density functions for speciation times can be obtained for any pure-birth process if the speciation rate function itself is integrable.
The probability density for the set of speciation times under any time-dependent pure-birth process starting with one initial species is
\begin{eqnarray}
f(\mathbb{T}|N(t_0)\!=\!1) & = & f(\mathbb{T}|N(t_0)\!=\!1,S(1,t_0,T)) \nonumber \\
& = & \exp\left(\int_{t_0}^T\lambda(x)dx\right)  \times\prod_{i=1}^{n-1}\Bigg(i\times\lambda(t_i)\times \exp\left(\int_{t_i}^T\lambda(x)dx\right)\Bigg)
\end{eqnarray}
or when the process starts with two initial species at time $t_1$
\begin{eqnarray}
f(\mathbb{T}|N(t_1)\!=\!2) & = & f(\mathbb{T}|N(t_1)\!=\!2,S(1,t_1,T)) \nonumber \\
 & = & \left[\exp\left(\int_{t_1}^T\lambda(x)dx\right)\right]^{2}  \times\prod_{i=2}^{n-1}\Bigg(i\times\lambda(t_i)\times \exp\left(\int_{t_i}^T\lambda(x)dx\right)\Bigg) \mbox{ .}
\end{eqnarray}
I present an example of non-constant pure-birth process, the decreasing rate pure-birth process, in the appendix.

\leveltwo{The pure-death process}
The pure-death process is defined by a speciation rate $\lambda(t) = 0$ and any extinction rate $\mu(t) > 0$. 
Therefore, the number of species is monotonically decreasing and the probability of any reconstructed tree is zero because speciation events cannot occur. 
Nevertheless, it may still be of interest to compute the survival probability of one species and the probability and time of extinction of $n$ species \citep{Nee2006}.

The diversification rate integral simplifies to $r(s,t) = \int_s^t \mu(x)dx$ in the general case. 
Then, the probability of survival, or no extinction, is 
\begin{eqnarray}
P(N(T)\!>\!0|N(t)\!=\!1) & = & \left(1+\int\limits_t^{T} \bigg(\mu(s) \exp(r(t,s))\bigg) ds\right)^{-1} \nonumber \\
& = & \exp\left(-\int_t^T \mu(x)dx\right)
\end{eqnarray} 
which is equivalent to the probability that no event occurs until time $T$ of an exponentially distributed random variable with rate $\mu(t)$. 

The probability of extinction of all $n$ species is
\begin{eqnarray}
P(N(T)\!=\!0|N(t)\!=\!n) & = & \prod_{i=1}^n \left[1-\exp\left(-\int_t^T \mu(x)dx\right)\right]
\end{eqnarray} 
which is obtained by using the fact that the extinction times of each species is \emph{iid} from an exponential distribution with rate $\mu(t)$.

\leveltwo{The time-dependent critical-branching process}
The critical-branching process considers the scenario in which speciation and extinction rates are equal: $\lambda(t) = \mu(t)$. 
The immediate consequence is that the diversification rate integral equals zero ($r(t,T) = \int_t^T\mu(s)-\lambda(s)ds = 0$). 
The probability density that at least one lineage survives is
\begin{eqnarray}
P(N(T)\!>\!0|N(t)\!=\!1) & = & \left(1+\int\limits_t^{T} \mu(s) ds\right)^{-1} \mbox{ .}
\end{eqnarray}

The critical-branching process has some interesting properties when conditioned on survival of at least on lineage. 
The expected number of lineage after time $T$ is 
\begin{eqnarray}
E[N(T) | S(1,t_0,T)] & = & \big(P(N(T)\!>\!0|N(t_0)\!=\!1)\exp(r(t_0,T))\big)^{-1} \nonumber \\
 & = & 1+\int\limits_t^{T} \mu(s) ds \mbox{ .}
\end{eqnarray}
Hence, the expected number of species increases towards infinity even under the critical-branching process. However, if the process is not conditioned on survival, the expected number of species is
\begin{eqnarray}
E[N(T)] & = & \big(\exp(r(t,T))\big)^{-1} \nonumber \\
& = & 1 
\end{eqnarray}
which means that every species alive at time $t$ is expected to leave exactly one descendant species at time $T$.

The small difference---whether we condition on the survival---has the rather large impact that the expected number of species monotonically increases even under a critical-branching process. 
Furthermore, the same effect can be observed when inferring diversification rates: the inferred extinction rate is larger when we condition on the survival of the process. 
Hence, if only a single reconstructed tree is analyzed and the diversification rates are estimated for this single tree, then it may be preferable to not condition on survival. 
Only if many trees are used together to estimate the diversification rates the condition on survival is justified because all trees have necessarily in common that the process did not result into extinction.

\leveltwo{The birth-death process is not scale invariant}
It is tempting to assume that the birth-death process is scale invariant; that is, that the probability of the reconstructed tree is the same under a given setting and when the tree and diversification rates are scaled (\EG from units in million years to units in 100 million years). 
Unfortunately, the assumption is wrong. 
To clarify this consider a constant-rate pure-birth process with rate $b$ and ages $\mathbb{A}$. 
Now scale the ages so that $\mathbb{A} = c\mathbb{A}'$ and the speciation rate $b = b'/c$. T
he resulting probability densities are
\begin{eqnarray}
n! b^{n-1} \times \exp\left(b \sum_{i=0}^{n-1}a_i)\right) & \neq & n! (b')^{n-1} \times \exp\left(b' \sum_{i=0}^{n-1}a_i')\right) \nonumber \\
n! b^{n-1} \times \exp\left(b \sum_{i=0}^{n-1}a_i)\right) & \neq & n! \left(\frac{b}{c}\right)^{n-1} \times \exp\left(\frac{b}{c} \sum_{i=0}^{n-1}a_i c)\right) \nonumber \\
1  & \neq & \left(\frac{1}{c}\right)^{n-1}
\end{eqnarray}
This holds under any time-dependent birth-death process.

The issue concerning the scale invariance applies only to the probability density (and the likelihood). 
Maximum likelihood estimates of diversification rates on the other hand are not effected because the difference between the scaled and unscaled probability density is a fixed factor ($(\frac{1}{c})^{n-1}$). 
However, the problem may effect joint inference of the reconstructed tree and diversification rates because a short tree with high diversification rates will give a higher probability than a long tree with low diversification rates.

\leveltwo{Conclusions}

In the present paper I have presented an approach to derive the probability density of the observed speciation times of a reconstructed tree, or the reconstructed tree itself, for any time-dependent birth-death process including mass-extinction events and uniform taxon sampling. 
I have demonstrated the use of the approach by deriving the analytical solutions of the probability density under the birth-death-shift process with multiple mass-extinction events. 
In the appendix I provide the probability density functions of other commonly used (time-dependent) birth-death process.

These models can be used for likelihood inference without the need of numerical integration inside the likelihood function. 
Computing the likelihood is fast and thus can be used even in computationally demanding methods such as Bayesian inference using Markov chain Monte Carlo sampling.

The common notation used here unifies all of the commonly used variants of the time-dependent birth-death process. 
Additionally, I showed how to convert the probability density function, \EG if the data are reconstructed trees or set of divergence times.
My hope is that this compendium facilitates comparisons among different birth-death processes and simplifies their application.

\section*{Acknowledgements}
I would like to thank Brian Moore and Fredrik Ronquist for comments on the manuscript.

\bibliographystyle{apalike}
\bibliography{literature}

\newpage
\noindent{\LARGE\bf Appendix I}
\setcounter{section}{0}
\renewcommand{\thesection}{\Alph{section}}
\levelone{Pure birth processes}

\leveltwo{Constant speciation rate}
The constant-rate pure birth process has rate $\lambda(t) = b$ and is arguably the simplest birth-death process. It is therefore often used as a null-model \citep{Yule1925,Nee2006}.
The rate integral is obtained by $r(t,s) = \int_t^s(\mu(x) - \lambda(x)) dx = b(s-t)$.
The probabilities for the number of species is given by
\begin{eqnarray}
P(N(T)\!=\!n|N(t)\!=\!1) & = & (1-P(N(T)\!>\!0|N(t)\!=\!1)\exp(r(t,T)))^{n-1} \nonumber\\
& & \times P(N(T)\!>\!0|N(t)\!=\!1)^2 \exp(r(t,T))  \nonumber \\
& = & (1-\exp(b(T-t)))^{n-1} \times \exp(b(T-t)) \mbox{ .}
\end{eqnarray}
The probability density of the speciation times for the process starting with one initial species at time $t_0$ is then given by
\begin{eqnarray}
f(\mathbb{T}|N(t_0)\!=\!1)  & = & \exp(b(T-t_0)) \times\prod_{i=1}^{n-1}\big(i\times b \times \exp(b(T-t_i))\big) \label{eq:constPureBirth}
\end{eqnarray}
and for the process starting with two initial species at time $t_1$ is
\begin{eqnarray}
f(\mathbb{T}|N(t_1)\!=\!2)  & = & \left[\exp(b(T-t_1))\right]^2 \times\prod_{i=2}^{n-1}\big(i\times b \times \exp(b(T-t_i))\big) \mbox{ .}
\end{eqnarray}
Note that it is common to use the age of the speciation events instead, where $a_i = T - t_i$, and to assume that $t_0 = 0$ and thus $a_0 = T$, which gives
\begin{eqnarray}
f(\mathbb{A}|N(t_0)\!=\!1)  & = & n! b^{n-1} \times \exp\left(b \sum_{i=0}^{n-1}a_i) \right) \mbox{ .} \label{eq:constPureBirthAges}
\end{eqnarray}
Equation~(\ref{eq:constPureBirthAges}) corresponds to Equation~(9) in \cite{Rannala1996} although Rannala and Yang computed the probability of a specific labeled history and conditioned on obtaining $n$ species  -- for the conversion see Equation~(\ref{eq:conversion}) and Equation~(\ref{eq:timesN}).

\levelthree{With uniform taxon sampling}
The constant-rate pure birth process can be extended to include uniform taxon sampling. However, the probability of survival does not equal one anymore because of the possibility to sample zero species.
Therefore, the probability of survival is given by
\begin{eqnarray}
\lefteqn{P(N(T)\!>\!0|N(t)\!=\!1)} \nonumber\\
& = & \left(1+\int\limits_t^{T} \bigg(\mu'(s) \exp(r'(t,T))\bigg) ds - \frac{\rho-1}{\rho} \exp(r'(t,s))\right)^{-1} \nonumber \\
& = & \left(1 - \frac{\rho-1}{\rho} \exp(b(T-t))\right)^{-1}
\mbox{ .} \label{eq:constPureBirthUniformSampling}
\end{eqnarray}
Furthermore, the probability of the set of speciation times for the process starting with one initial species is
\begin{eqnarray}
f(\mathbb{T}|N(t_0)\!=\!1)  & = & \frac{\rho \exp(b(T-t_0))}{\left(\rho - (\rho-1) \exp(b(T-t_0))\right)^{2}} \nonumber \\
&    & \times\prod_{i=1}^{n-1}\left(i\times b \times \frac{\rho \exp(b(T-t_i))}{\left(\rho - (\rho-1) \exp(b(T-t_i))\right)^{2}}\right)
\end{eqnarray}
and for the process starting with two initial species at time $t_1$ is
\begin{eqnarray}
f(\mathbb{T}|N(t_1)\!=\!2)  & = & \left[\frac{\rho \exp(b(T-t_1))}{\left(\rho - (\rho-1) \exp(b(T-t_1))\right)^{2}}\right]^2 \nonumber \\
&    & \times\prod_{i=2}^{n-1}\left(i\times b \times \frac{\rho \exp(b(T-t_i))}{\left(\rho - (\rho-1) \exp(b(T-t_i))\right)^{2}}\right) \mbox{ .}
\end{eqnarray}

\levelthree{Including a single mass-extinction event}
Similarly, the constant-rate pure birth process can be extended to include a single mass-extinction event at time $t_m$ with survival probability $\rho_m$.
The probability of survival is given by
\begin{eqnarray}
\lefteqn{P(N(T)\!>\!0|N(t)\!=\!1)} \nonumber\\
& = & \left(1+\int\limits_t^{T} \bigg(\mu'(s) \exp(r(t,s))\bigg) ds - \frac{\rho_m-1}{\rho_m} \exp(r'(t,t_m))\right)^{-1} \nonumber \\
& = & \left(1 - \frac{\rho_m-1}{\rho_m} \exp(b(t_m-t)))\right)^{-1}
\mbox{ .} \label{eq:constPureBirthMassExtinction}
\end{eqnarray}
The probability of the set of speciation times for the process starting with a single species at time $t_0$ is then
\begin{eqnarray}
f(\mathbb{T}|N(t_0)\!=\!1)  & = & \frac{(\rho_m-1) \exp(b(T-t_0))}{\left(1 - \frac{\rho_m-1}{\rho_m} \exp(b(t_m-t_0))\right)^{2}} \nonumber \\
&    & \times\prod_{i=1}^{n-1} i\times b \times 
\begin{cases} \exp(b(T-t_i)) & \text{ if } t_m < t_i \\
\frac{(\rho_m-1) \exp(b(T-t_i))}{\left(1 - \frac{\rho_m-1}{\rho_m} \exp(b(t_m-t_i))\right)^{2}}  & \text{ otherwise} \nonumber
\end{cases}\\
\end{eqnarray}
and, again, for the process starting with two initial species at time $t_1$ is
\begin{eqnarray}
f(\mathbb{T}|N(t_1)\!=\!2)  & = & \left[\frac{(\rho_m-1) \exp(b(T-t_1))}{\left(1 - \frac{\rho_m-1}{\rho_m} \exp(b(t_m-t_1))\right)^{2}}\right]^2 \nonumber \\
&    & \times\prod_{i=2}^{n-1} i\times b \times 
\begin{cases} \exp(b(T-t_i)) & \text{ if } t_m < t_i \\
\frac{(\rho_m-1) \exp(b(T-t_i))}{\left(1 - \frac{\rho_m-1}{\rho_m} \exp(b(t_m-t_i))\right)^{2}}  & \text{ otherwise.} \nonumber
\end{cases}\\
\end{eqnarray}

\levelthree{Including multiple mass-extinction events}

Similarly to the previous equation, multiple mass-extinction events can be included. I use again the notation that $m_k$ denotes the time and $\rho_k$ the survival probability of the $k^{th}$ mass-extinction event. The probability of survival is
\begin{eqnarray}
\lefteqn{P(N(T)\!>\!0|N(t)\!=\!1)} \nonumber\\
& = & \left(1 - \sum_{k = 0}^j\frac{\rho_k-1}{\prod_{l=0}^k\rho_l} \exp\bigg(b(m_k-t)\bigg)\right)^{-1} \mbox{ .} \label{eq:constPureBirthMassExtinction}
\end{eqnarray}
This probability density can then be inserted to compute the probability density of a reconstructed tree (or the set of speciation times).

\leveltwo{Exponentially decaying speciation rate}
Adaptive or rapid radiation are represented by an exponentially decaying speciation rate \citep{Rabosky2006,Rabosky2008,Morlon2011,Hohna2013b}. The common approach uses the speciation rate function $\lambda(t) = \lambda_0 \exp(-\alpha t)$. Thus, the rate integral is $r(s,t) = \int_s^t(\mu(x) - \lambda(x)) dx = \frac{\lambda_0}{\alpha}(\exp(-\alpha s)-\exp(-\alpha t))$. The probability density of a reconstructed tree cannot be simplified any further but still can be computed analytically by inserting the speciation rate and the diversification rate integral into Equation~(\ref{eq:times}), which yields for the process starting with a single species at time $t_0$
\begin{eqnarray}
f(\mathbb{T}|N(t_0)\!=\!1)  & = & \exp\left( \frac{\lambda_0}{\alpha}\Big(\exp(-\alpha t_0) - \exp(-\alpha T)\Big) \right) \nonumber \\
& & \times\prod_{i=1}^{n-1}\left(i\times \lambda_0 \exp(-\alpha t_i) \times \exp\bigg( \frac{\lambda_0}{\alpha}\Big(\exp(-\alpha t_i) - \exp(-\alpha T)\Big) \bigg) \right) \nonumber \\
\end{eqnarray}
and for the process starting with two species at time $t_1$
\begin{eqnarray}
f(\mathbb{T}|N(t_1)\!=\!2)  & = & \left[\exp\left( \frac{\lambda_0}{\alpha}\Big(\exp(-\alpha t_1) - \exp(-\alpha T)\Big) \right) \right]^2 \nonumber \\
& & \times\prod_{i=2}^{n-1}\left(i\times \lambda_0 \exp(-\alpha t_i) \times \exp\bigg( \frac{\lambda_0}{\alpha}\Big(\exp(-\alpha t_i) - \exp(-\alpha T)\Big) \bigg) \right) \mbox{ .} \nonumber \\
\end{eqnarray}

\levelone{The constant-rate Birth-Death Process}

In the previous sections I derived the probability density of survival of a pure birth process and a pure death process, thus either the speciation or the extinction rate was set to zero. This setting may be of interest in a purely theoretical study or when studying boundary conditions, but it is not biologically realistic. In this section I will elaborate on the probability densities under a constant-rate birth-death process.

The constant speciation and extinction rates are: $\lambda(t) = b$ and $\mu(t) = d$. This yields the diversification rate integral
\begin{eqnarray}
r(t,s) & = & (d - b) \times (s-t) \mbox{ .}
\end{eqnarray}
Then, the probability of survival and the probability of $n$ species at the present time are respectively
\begin{eqnarray}
P(N(T)\!>\!0|N(t)\!=\!1) & = & \frac{b - d}{b - d \text{e}^{-(b-d)(T-t)}}\\
P(N(T)\!=\!n|N(t)\!=\!1) & = & \left(\frac{b}{d}\right)^{n-1}P(N(T)\!=\!1|N(t)\!=\!1) \nonumber \\
 & & \times [P(N(T)\!=\!0|N(t)\!=\!1)]^{n-1} \mbox{ .}
\end{eqnarray}
The probability of extinction of the process and the probability of obtaining exactly one species are often needed in several of the following probability density functions -- such as the probability density of the set of speciation times. Hence, I provide these probabilities for convenience here: 
\begin{eqnarray}
P(N(T)\!=\!0|N(t)\!=\!1) & = & 1 - P(N(T)\!>\!0|N(t)\!=\!1) \nonumber \\
                                      & = & \frac{d(1-\text{e}^{-(b-d)(T-t)})}{b - d\text{e}^{-(b-d)(T-t)}}\\
P(N(T)\!=\!1|N(t)\!=\!1) & = & \frac{(b-d)^2\text{e}^{-(b-d)(T-t)}}{(b - d\text{e}^{-(b-d)(T-t)})^2} \mbox{ .}
\end{eqnarray}
Additionally, the expression $1- P(N(T)\!>\!0|N(t)\!=\!1)\exp(r(t,T))$ will be needed:
\begin{eqnarray}
1- P(N(T)\!>\!0|N(t)\!=\!1)\exp(r(t,T))  & = & 1 - \frac{(b - d)\text{e}^{-(b-d)(T-t)}}{b - d \text{e}^{-(b-d)(T-t)}} \nonumber \\
  & = &  \frac{b(1-\text{e}^{-(b-d)(T-t)})}{b - d\text{e}^{-(b-d)(T-t)}} \nonumber \\
 & = & \frac{b}{d} P(N(T)\!=\!0|N(t)\!=\!1)  \mbox{ .}
\end{eqnarray}

The probability distribution function and the probability density function of a speciation event in the reconstructed tree are derived by using the above equations:
\begin{eqnarray}
F(t|t_0\leq t \leq T) & = & 1- \frac{1-P(N(T)\!>\!0|N(t)\!=\!1)\exp(r(t,T))}{1-P(N(T)\!>\!0|N(t_0)\!=\!1)\exp(r(t_0,T))} \nonumber \\
& = & 1 - \frac{P(N(T)\!=\!0|N(t)\!=\!1)}{P(N(T)\!=\!0|N(t_0)\!=\!1)} \nonumber \\
& = & 1 - \frac{1-\text{e}^{-(b-d)(T-t)}}{b - d\text{e}^{-(b-d)(T-t)}} \times \frac{b - d\text{e}^{-(b-d)(T-t_0)}}{1-\text{e}^{-(b-d)(T-t_0)}} \label{eq:constBDPdistTime}\\
f(t|t_0\leq t \leq T) & = & \frac{\lambda(t) P(N(T)\!=\!1|N(t)\!=\!1)}{1 - P(N(T)\!>\!0|N(t_0)\!=\!1)\exp(r(t_0,T))}  \nonumber \\
 & = & d\frac{P(N(T)\!=\!1|N(t)\!=\!1)}{P(N(T)\!=\!0|N(t_0)\!=\!1)}  \nonumber \\
 & = & d \frac{(b-d)^2\text{e}^{-(b-d)(T-t)}}{(b - d\text{e}^{-(b-d)(T-t)})^2} \times \frac{b - d\text{e}^{-(b-d)(T-t_0)}}{1-\text{e}^{-(b-d)(T-t_0)}}  \label{eq:constBDPdensTime}
\end{eqnarray}
Equation~(\ref{eq:constBDPdistTime}) and Equation~(\ref{eq:constBDPdensTime}) correspond to Equation~(1) and (2) in \cite{Hohna2011}. Note that in \cite{Hohna2011} we considered time going backwards into the past and therefore the distribution function is slightly modified (it is $1 - F(t|t_0\leq t \leq T)$). Time going backwards leads to same equation when the diversification rates are constant and may simplify some equations (see Equation~(\ref{eq:constPureBirthAges})). However, if the diversification rates vary over time, then the direction of time has to match.

\leveltwo{The probability density of the set of speciation times}
The probability density function of the set of speciation times of a reconstructed tree under the constant-rate birth-death process is given by
\begin{eqnarray}
f(\mathbb{T}|N(t_0)\!=\!1)  & = & (n-1)! b^{n-1} \frac{(b-d)^3\text{e}^{-(b-d)(T-t_0)}}{(b - d\text{e}^{-(b-d)(T-t_0)})^3}  \nonumber\\
& & \times\prod_{i=1}^{n-1}\left(\frac{(b-d)^2\text{e}^{-(b-d)(T-t_i)}}{(b - d\text{e}^{-(b-d)(T-t_i)})^2}\right) \label{eq:constBirthDeathTimes} \mbox{ .}
\end{eqnarray}
By conditioning on survival of the process and starting with two species at the time of the most recent common ancestor ($t_{MRCA} = t_1$) I obtain
\begin{eqnarray}
f(\mathbb{T}|N(t_1)\!=\!2,S(2,t_1,T))  & = & (n-1)! b^{n-2} \left(\frac{(b-d)^2\text{e}^{-(b-d)(T-t_1)}}{(b - d\text{e}^{-(b-d)(T-t_1)})^2}\right)^{2}  \nonumber\\
& & \times\prod_{i=2}^{n-1}\left(\frac{(b-d)^2\text{e}^{-(b-d)(T-t_i)}}{(b - d\text{e}^{-(b-d)(T-t_i)})^2}\right) \label{eq:constBirthDeathTimesMRCA}
\end{eqnarray}
which is equivalent to Equation~(20) in \cite{Nee1994}. Equation~(\ref{eq:constBirthDeathTimesMRCA}) can be used to infer the speciation and extinction rate under a constant-rate birth-death process when the tree was reconstructed from molecular data.

\leveltwo{With uniform taxon sampling}
Now, I extend the constant-rate birth-death process to include uniform taxon sampling \citep{Nee1994,Yang1997,Stadler2009,Hohna2011,Hohna2013b}. First, I provide the probability density of survival of at least one lineage:
\begin{eqnarray}
\lefteqn{P(N(T)\!>\!0|N(t)\!=\!1)} \nonumber\\
& = & \left(1+\int\limits_t^{T} \bigg(\mu'(s) \exp(r'(t,T))\bigg) ds - \frac{\rho-1}{\rho} \exp(r'(t,T))\right)^{-1} \nonumber \\
& = & \left(1+\frac{d}{(d-b)} \bigg(\text{e}^{(d-b)(T-t)} - 1\bigg) - \frac{\rho-1}{\rho} \text{e}^{(d-b)(T-t)}\right)^{-1} \nonumber \\
& = & \frac{\rho(b-d)}{\rho b + \bigg(b(1-\rho) - d\bigg) \text{e}^{(d-b)(T-t)}}
 \label{eq:constBirthDeathUniformSampling}
\end{eqnarray}
which corresponds to Equation~(1) in \cite{Yang1997} with the assumption that $t=0$.

The probability density of the set of speciation times is
\begin{eqnarray}
f(\mathbb{T}|N(t_0)\!=\!1)  & = & (n-1)! b^{n-1} \frac{\text{e}^{-(b-d)(T-t_0)}}{\rho}  \nonumber\\
& & \times \left(\frac{\rho(b-d)}{\rho b + \bigg(b(1-\rho) - d\bigg) \text{e}^{(d-b)(T-t_0)}}\right)^3 \nonumber \\
& & \times\prod_{i=1}^{n-1}\left(\frac{\rho(b-d)^2\text{e}^{-(b-d)(T-t_i)}}{\bigg(\rho b + \big(b(1-\rho) - d\big)\text{e}^{-(b-d)(T-t_i)}\bigg)^2}\right) \label{eq:constBirthDeathTimes}
\end{eqnarray}
or if conditioned on starting with two species at $t_1$ and both survive until the present, then the probability density is
\begin{eqnarray}
f(\mathbb{T}|N(t_1)\!=\!2,S(2,t_1,T))  & = & (n-1)! b^{n-2} \left(\frac{(b-d)\text{e}^{-(b-d)(T-t_1)}}{\rho b + \bigg(b(1-\rho) - d\bigg) \text{e}^{(d-b)(T-t_1)}}\right)^2 \nonumber\\
& & \times\prod_{i=2}^{n-1}\left(\frac{\rho(b-d)^2\text{e}^{-(b-d)(T-t_i)}}{\bigg(\rho b + \big(b(1-\rho) - d\big)\text{e}^{-(b-d)(T-t_i)}\bigg)^2}\right) \label{eq:constBirthDeathTimes} \mbox{ .}
\end{eqnarray}

\leveltwo{The birth-death process with constant rates and mass-extinction events}

\levelthree{A single mass-extinction event}
Let the time of mass-extinction event be denoted by $t_m$ and the mass-extinction survival probability $\rho_m$.
The probability of survival of the process can be computed by
\begin{eqnarray}
\lefteqn{P(N(T)\!>\!0|N(t)\!=\!1)} \nonumber \\
& = & \left(1+\int\limits_t^{T} \bigg(\mu(s) \exp(r(t,s))\bigg) ds\right)^{-1} \nonumber \\
& = & \Bigg(1+\int\limits_t^{t_m-\Delta t} \bigg(d \text{e}^{(d-b)(s-t)}\bigg) ds + \int\limits_{t_m-\Delta t}^{t_m+\Delta t} \bigg(\mu(s) \text{e}^{r(t,s)}\bigg) ds \nonumber \\
& & + \int\limits_{t_m+\Delta t}^{T} \bigg(d \text{e}^{(d-b)(s-t)-\ln(\rho_m)}\bigg) ds\Bigg)^{-1} \nonumber \\
& \overset{\Delta t \to 0}{=} & \Bigg(1+\frac{d}{d-b}\left( \text{e}^{(d-b)(t_m-t)} - 1 \right) - \frac{\rho_m-1}{\rho_m} \text{e}^{(d-b)(t_m-t)} \nonumber \\
& &  +\frac{d}{\rho(d-b)}\left( \text{e}^{(d-b)(T-t)} - \text{e}^{(d-b)(t_m-t)} \right)\Bigg)^{-1} \mbox{ .} \label{eq:constBirthDeathMassExtinctionSurvival}
\end{eqnarray}
Using Equation~(\ref{eq:constBirthDeathMassExtinctionSurvival}) it is possible to compute the probability of the set of speciation times but I will omit the equation here because the resulting equation looks messy and its derivation should be clear from the previous sections.

\levelthree{Multiple mass-extinction events}
Finally, I derive the probability density of the speciation times of a reconstructed tree under a constant-rate birth-death process with multiple mass-extinction events. As before, the times of the mass-extinction events are denoted by $m_k$ and the survival probability by $\rho_k$. The probability of at least one surviving lineage at time $T$ is
\begin{eqnarray}
\lefteqn{P(N(T)\!>\!0|N(t)\!=\!1)} \nonumber \\
& = & \Bigg(1+ \sum\limits_{k = i}^{j} \left( \frac{d}{(d-b)\prod_{l=i}^{k-1}\rho_l}\big(\text{e}^{(d-b)(m_k-t)} - \text{e}^{(d-b)(m_{k-1}-t)} \big) - \frac{\rho_k-1}{\prod_{l=i}^{k}\rho_l} \text{e}^{(d-b)(m_k-t)}\right) \Bigg)^{-1} \label{eq:constBirthDeathMultiMassExtinctionSurvival}
\end{eqnarray}
conditioned on starting with one lineage at time $t$.

\end{document}